# Virtual Cylindrical PET for Efficient DOI Image Reconstruction with Sub-millimetre Resolution

Francisco E. Enríquez-Mier-y-Terán[a,b,*], Andre Z. Kyme[a,b], Georgios Angelis[a,b,c], Steven R. Meikle[b,c,d],

[a]School of Biomedical Engineering, Faculty of Engineering, The University of Sydney, Sydney, NSW 2006, Australia
[b]Brain and Mind Centre, The University of Sydney, Sydney, NSW 2050, Australia
[c]Sydney Imaging Core Research Facility, The University of Sydney, NSW 2050, Australia
[d]School of Health Sciences, Faculty of Medicine and Health, The University of Sydney, Sydney, NSW 2050, Australia

*Corresponding author.
E-mail address: fenr7890@uni.sydney.edu.au (F.E. Enríquez-Mier-y-Terán)

## Abstract

Objective:
Image reconstruction in high resolution, narrow bore PET scanners with depth of interaction (DOI) capability presents a substantial computational challenge due to the very high sampling in detector and image space. The aim of this study is to evaluate the use of a virtual cylinder in reducing the number of lines of response (LOR) for DOI-based reconstruction in high resolution PET systems while maintaining uniform sub-millimetre spatial resolution.

Approach:
Virtual geometry was investigated using the awake animal mousePET as a high resolution test case. Using GATE, we simulated the physical scanner and three virtual cylinder implementations with detector size 0.7405 mm, 0.4712 mm and 0.3575 mm (vPET1, vPET2 and vPET3, respectively). The virtual cylinder condenses physical LORs stemming from various crystal pairs and DOI combinations, and which intersect a single virtual detector pair, into a single virtual LOR. Quantitative comparisons of the point spread function (PSF) at various positions within the field of view (FOV) were compared for reconstructions based on the vPET implementations and the physical scanner. We also assessed the impact of the anisotropic PSFs by reconstructing images of a micro Derenzo phantom.

Main results:
All virtual cylinder implementations achieved LOR data compression of at least 50% for DOI PET reconstruction. PSF anisotropy in radial and tangential profiles was chiefly influenced by DOI resolution and only marginally by virtual detector size. Spatial degradation introduced by virtual cylinders was most prominent in the axial profile. All virtual cylinders achieved sub-millimetre volumetric resolution across the FOV when 6-bin DOI reconstructions (3.3 mm DOI resolution) were performed. Using vPET2 with 6 DOI bins yielded nearly identical reconstructions to the non-virtual case in the transaxial plane, with an LOR compression ratio of 86%. Resolution modelling significantly reduced the effects of the asymmetric PSF arising from the non-cylindrical geometry of mousePET.

<u>Significance:</u>

Narrow bore and high resolution PET scanners require detectors with DOI capability, leading to computationally demanding reconstructions due to the large number of LORs. In this study, we show that DOI PET reconstruction with 50-86% LOR compression is possible using virtual cylinders while maintaining sub-millimetre spatial resolution throughout the FOV. The methodology and analysis can be extended to other scanners with DOI capability intended for high resolution PET imaging.

**Introduction**

Narrow bore PET scanners, typical of small animal and organ-dedicated systems, aim to achieve uniformly high spatial resolution and sensitivity throughout the field of view (FOV). For high sensitivity, scanners typically consist of detectors comprising long crystals with minimal gaps between them (Tashima and Yamaya 2016, Carson et al 2021, Wang et al 2022). For high resolution, detectors often use finely segmented (thin) crystals (Zeng et al 2023, Kang et al 2023, Kuang et al 2023). However, the combination of narrow bore and long crystals means that annihilation photons impinging at oblique angles with respect to the crystal front face are commonly mispositioned (known as parallax error), resulting in degraded and non-uniform spatial resolution. To correct this, detectors must account for the depth of interaction (DOI) of the annihilation photons. In DOI-capable detectors, the lines of response (LORs) are not assigned to a single crystal location (e.g., the crystal front face) but to the centre of the DOI bin where the interaction took place. More precise DOI localisation of events leads to reduction of parallax errors and more uniform spatial resolution across the FOV (Kyme et al 2017, Kang et al 2021, Zeng et al 2023).

An important cost of DOI-capable systems is that the number of LORs increases $\sim N^2$ ($N$ the number of DOI bins). For iterative reconstruction algorithms, this can quickly create computational bottlenecks from the forward and back projection at each iteration, and calculation of the system matrix. The latter requires passing through all possible LORs within the system and therefore elements of the system matrix are often pre-calculated once and stored on disk (Zhou and Qi 2011). A one-off system matrix calculation, albeit a lengthy one, is not always possible though. For example, LOR-based motion compensation with precise attenuation correction requires repeated computation that accounts for the time-weighted effect of each sampled movement of the subject (Rahmim et al 2004, Angelis et al 2014). This is not always practical, especially for DOI-capable systems.

As the limits of PET spatial resolution continue to be pushed in novel application-specific scanners with DOI and motion compensation, there is a parallel need for computationally efficient reconstruction algorithms to support this trend. One approach for LOR reduction (compression) is the use of virtual geometries to rebin multiple physical LORs from different crystals and DOI bins into a single virtual LOR (Fig. 1) (Li et al 2015, Zhang et al 2016, Groll et al 2017, Kim et al 2018, Wang et al 2022). Previous work on virtual geometries for PET reconstruction is, however, rather inconclusive for several reasons: (i) performance comparisons between uncompressed (full) and compressed system matrices are not always reported (Groll et al 2017, Wang et al 2022) or have been inconclusive (Li et al 2015); (ii) conclusions regarding their effect on preserving spatial resolution are inconsistent, with some suggesting it is well preserved (Zhang et al 2016) and others that it is degraded unless sub-sampling methods are employed (Kim et al 2018); and (iii) there has been very little reported on the use of a virtual cylinder with smaller radius than the physical scanner (Li et al 2015, Wang et al 2022). Compared to virtual geometries coinciding with the first DOI bin of the physical scanner (Zhang et al 2016, Groll et al 2017, Kim et al 2018), the use of a smaller virtual ring radius provides further LOR compression. This is at the expense of a reduced effective FOV and spatial resolution, but the precise cost and implications are not well understood.

In this work we implemented virtual cylindrical detectors for DOI PET image reconstruction in the context of the open-field Mouse Brain PET scanner (mousePET) (Kyme et al 2017, Enriquez-Mier-Y-Teran et al 2021), an overlapping box-shaped PET scanner for awake animal imaging. The implementation is based on a virtual cylinder configuration with smaller radius than the original FOV. The combination of (i) finely segmented crystals (0.785 mm × 0.785 mm × 20 mm) with 3-mm DOI resolution (Kyme et al 2017) and (ii) the need for a time-varying system matrix for motion correction (Rahmim et al 2004, Angelis et al 2014), makes the scanner a highly LOR-intensive PET system and, therefore, a good case study for investigating potential benefits of a virtual detector. The primary objective of this study was to assess the impact of the virtual cylinder compression method (referred to as 'virtual PET') on the spatial resolution and DOI performance of mousePET, and hence to extrapolate some general principles of virtual detector performance for other high resolution systems.

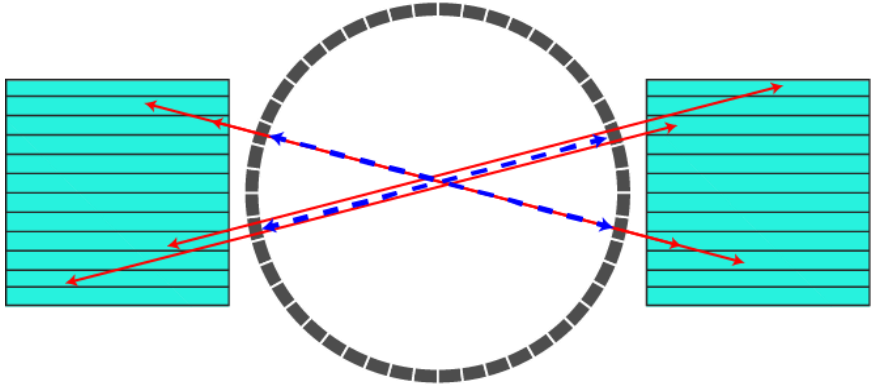

**Figure 1**. DOI-based PET reconstruction using a virtual cylindrical geometry. A virtual detector ring (black) consolidates multiple physical LORs (red lines) from various crystals and DOI bins in the physical detector (cyan) into single virtual LORs (dashed blue lines).

## 2. Methods

The experimental work is organized as follows: Section 2.1 outlines the process by which physical LORs are mapped onto virtual LORs; Section 2.2 details the simulation of the mousePET scanner; Section 2.3 introduces the reconstruction pipeline, including computation of normalisation factors and leveraging scanner symmetries for statistical robustness; and Section 2.4 describes the simulation experiments we conducted to characterise and compare physical mousePET and virtual PET implementations.

2.1 Virtual PET Implementation

The implementation of a virtual cylindrical PET scanner proceeds similarly to (Li et al 2015). The end points of a physical LOR connecting crystals $i$ and $j$, and any DOI bin combination (assuming the LOR endpoints are at the centre of each DOI bin), are defined using six spatial coordinates: $\boldsymbol{S_i} = (x_i, y_i, z_i)$ and $\boldsymbol{S_j} = (x_j, y_j, z_j)$ (Fig 2a). It is convenient to express the LOR as a line, given by:

$$\boldsymbol{l} = \boldsymbol{S_i} + t\left(\frac{\boldsymbol{S_j} - \boldsymbol{S_i}}{\|\boldsymbol{S_j} - \boldsymbol{S_i}\|}\right) \quad (1)$$

where $\boldsymbol{l}$ is any point on the LOR and its position along the LOR starting from $\boldsymbol{S_i}$ is parameterised by $t$. The point can be decomposed into its 3 spatial coordinates according to:

$$x = x_i + t\beta(x_j - x_i) \quad (2.1)$$
$$y = y_i + t\beta(y_j - y_i) \quad (2.2)$$
$$z = z_i + t\beta(z_j - z_i) \quad (2.3)$$

with $\beta = \frac{1}{\|\boldsymbol{S_j} - \boldsymbol{S_i}\|}$.

The intersection, $(x', y', z')$, of this LOR with an axially aligned virtual cylinder of radius $R$ satisfies:

$$x'^2 + y'^2 = R^2 \qquad (3)$$

Substituting equations (2.1) and (2.2) into (3) gives:

$$(x_i^2 + y_i^2) + 2t\beta(x_i\Delta x + y_i\Delta y) + t^2\beta^2(\Delta x^2 + \Delta y^2) - R^2 = 0 \qquad (4)$$

where $\Delta x = x_j - x_i$ and $\Delta y = y_j - y_i$.

The solution for $t$ in this quadratic equation is given by:

$$t_{\pm} = \frac{\beta^{-1}}{(\Delta x^2 + \Delta y^2)}\left(\pm\sqrt{R^2(\Delta x^2 + \Delta y^2) - (x_i\Delta y - y_i\Delta x)^2} - (x_i\Delta x + y_i\Delta y)\right) \qquad (5)$$

which provides the intersection points of the LOR with the virtual cylinder:

$$\boldsymbol{S'}_i = \boldsymbol{S}_i + \frac{\min(t_{\pm})}{\beta}(\boldsymbol{S}_j - \boldsymbol{S}_i) \qquad (6.1)$$

$$\boldsymbol{S'}_j = \boldsymbol{S}_i + \frac{\max(t_{\pm})}{\beta}(\boldsymbol{S}_j - \boldsymbol{S}_i) \qquad (6.2)$$

Since the starting point on the LOR is $\boldsymbol{S}_i$, the minimum value of the scaling factor ($\min(t_{\pm})$) gives the solution to $\boldsymbol{S'}_i$ and the maximum value gives the solution to $\boldsymbol{S'}_j$.

Once all LOR intersections with the virtual cylinder have been calculated, a segmentation of the cylinder allows the definition of virtual detectors. We chose virtual detectors with equal length and width (i.e., even segmentation along the radial and $z$ directions). Physical LORs intersecting the same virtual detector pair are consolidated into a single virtual LOR (Fig. 1). The number and size of the virtual detectors, along with the radius of the virtual ring, can all be adjusted for a given application. This is important because the granularity of the virtual detectors impacts the coarseness of sampling in image space, which in turn impacts the trade-off between computational efficiency and accuracy. The virtual PET radius was set to $R = 33$ mm to accommodate mice inside the FOV. We evaluated three virtual detector sizes ($d$) (Fig. 2a): 0.7405 mm (vPET1), 0.4712 mm (vPET2), and 0.3575 mm (vPET3).

To assess the impact of DOI resolution on mousePET performance, we considered three DOI resolution settings - no DOI, 5-mm DOI resolution (comprising 4 DOI bins), and 3-mm DOI resolution (comprising 6 DOI bins) – for the three virtual PET implementations. (Note that 3-mm DOI with 6 bins is the specification of the final mousePET system.) The LOR compression ratios resulting from these implementations are shown in Table 1.

**Table 1.** LOR compression ratios for the different virtual PET implementations and number of DOI bins used during reconstruction. The number of physical LORs in mousePET as a function of the number of DOI bins is included as a reference (top row).

|       | no DOI bins | 4 DOI bins | 6 DOI bins |
|-------|-------------|------------|------------|
|       | 281,851,200 | 3,203,541,778 | 7,211,211,922 |
| **vPET1** | 0.38 | 0.95 | 0.97 |
| **vPET2** | 0.05 | 0.73 | 0.86 |
| **vPET3** | 0.02 | 0.50 | 0.69 |

2.2 System Simulation

The mousePET scanner (Fig. 2b) features a box-shaped geometry for high sensitivity (Habte et al 2007, Kyme et al 2017) and consists of 4 overlapping detector panels of 4 × 4 detector modules, each comprising a 23 × 23 LYSO crystal array (0.785 mm × 0.785 mm × 20 mm, crystal pitch = 0.85 mm) read out on both sides for DOI encoding by a 6 × 6 array of 3-mm SiPMs. This system was simulated using GEANT4 Application for Tomographic Emission (GATE) v8.2 (Jan et al 2004, Sarrut et al 2021). Coincidences were allowed between any pair of detector panels and the energy and timing resolution were set to 18% (at 511 keV) and 2 ns, respectively. A coincidence time window of 6 ns was used in all experiments. DOI was implemented by reassigning LORs to the centre of their corresponding DOI bins.

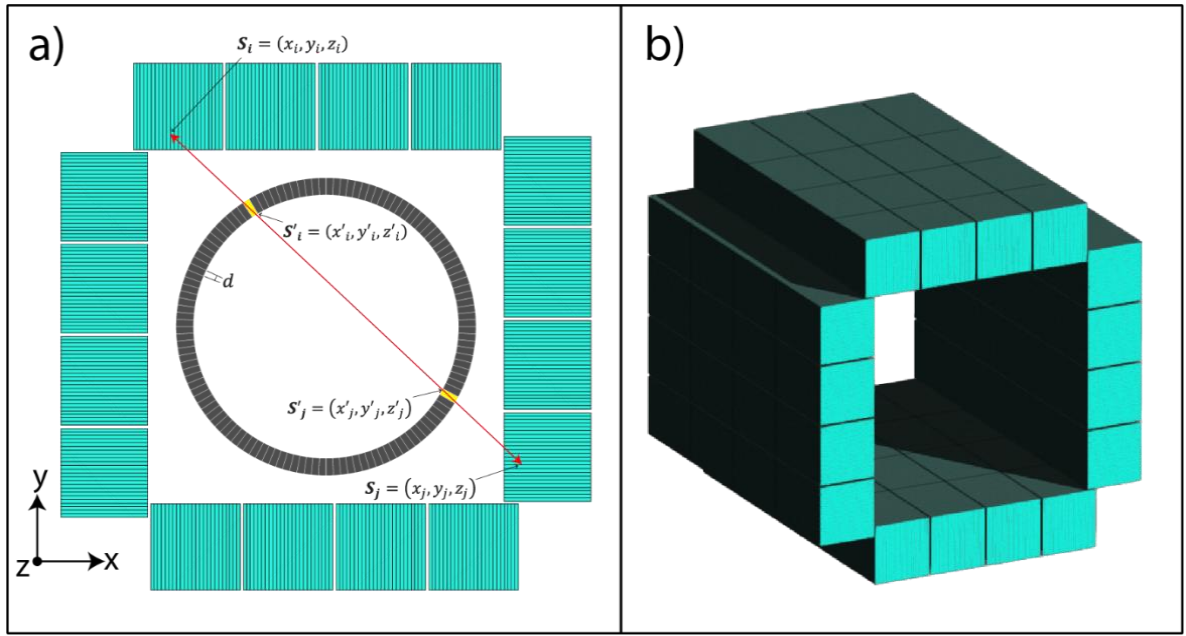

**Figure 2**. (**A**) The LOR (red line) between crystal $i$ and crystal $j$ is defined by its end points: $S_i = (x_i, y_i, z_i)$ and $S_j = (x_j, y_j, z_j)$. The LOR intersects the virtual ring at points $S'_i = (x'_i, y'_i, z'_i)$ and $S'_j = (x'_j, y'_j, z'_j)$. A single virtual LOR (not shown) connecting the centre of two virtual detectors with size $d$ (in yellow) replaces the red LOR and any other LOR intersecting the same virtual detector pair. (**B**) 3D rendering of the unconventional box-shaped mousePET scanner showing its 4 detector

panels. Each panel comprises 4 x 4 detector blocks, each consisting of 23 x 23 LYSO crystals read out from both ends for DOI capability.

2.3. Image Reconstruction

Image reconstruction was performed using a list-mode ordered subsets expectation maximisation (OSEM) algorithm (Reader et al 2002) with the sensitivity image pre-corrected for the different detector efficiencies:

$$x_q^{kS+s+1} = \frac{x_q^{kS+s}}{\sum_{p=1}^{P} n_p a_{pq}} \sum_{p \in T_s} a_{pq} \frac{1}{\sum_{q=1}^{Q} a_{pq} x_q^{kS+s}} \quad (7)$$

where $S$ is the number of subsets, $s \in \{1,\ldots,S\}$, $kS$ is the number of effective (full) iterations after $k$ passes through the data, $x_q^{kS+s}$ is the mean activity in voxel $q$ after the $kth$ iteration due to $T_s$ number of LORs in subset $s$, $a_{pq}$ is the probability of detecting an emission from voxel $q$ on LOR $p$, and $n_p$ is the normalisation factor of LOR $p$.

In all phantom experiments the image matrix size was set to $179 \times 179 \times 189$ voxels and the voxel size to $0.3925 \times 0.3925 \times 0.4250$ mm$^3$.

2.4 Normalisation

Normalisation factors $n_p$ were proportional to the detection efficiencies of the pair of detectors connecting the LOR. For LOR $p$, the normalisation coefficient is calculated as the ratio of the ideal measured counts ($A_p$) of a known object to the true measured counts ($M_p$), obtained experimentally or through Monte Carlo simulation:

$$n_p = \frac{M_p}{A_p} \quad (8)$$

To calculate $M_p$ values of both the mousePET scanner and virtual PET implementations, we conducted simulations using a uniform cylindrical source with an activity of 16 MBq. This source effectively covered the entire FOV with a radius of 32.8 mm. For the non-DOI and 4-bin DOI case we simulated acquisitions of 6 h duration to accurately estimate the correction factors. The simulation duration was extended to 10 h for the 6-bin DOI case. To expedite the simulation process we used back-to-back gamma events instead of a positron-emitting source. The $A_p$ values were calculated by forward projecting the cylindrical source through all possible LORs.

To better estimate the normalisation factors, we harnessed symmetries inherent in the scanner geometry, effectively augmenting the number of counts per LOR ($M_p$). In the transaxial plane, there are three rotational symmetries for the majority of the LORs

(Fig. 3a). Four more symmetries are obtained by mirroring the LOR across the axial plane and considering its corresponding three rotational symmetries (Fig. 3b). The exception to these symmetries is the small group of LORs shown in Figure 3c, which exhibit only three symmetries. The 4- and 8-fold increment in the number of counts per LOR was applied prior to calculating the normalisation coefficients of each system.

To normalize the virtual PET implementations, the correction term for a virtual LOR $p'$ ($n_{p'}$) was derived according to:

$$n_{p'} = \frac{\sum_{p \in p'} M_p}{A_{p'}} \qquad (3)$$

where the numerator represents the sum of $M_p$ terms associated with all physical LORs assigned to the virtual LOR, and the denominator $A_{p'}$ represents the forward projection along the virtual LORs for the same cylindrical source.

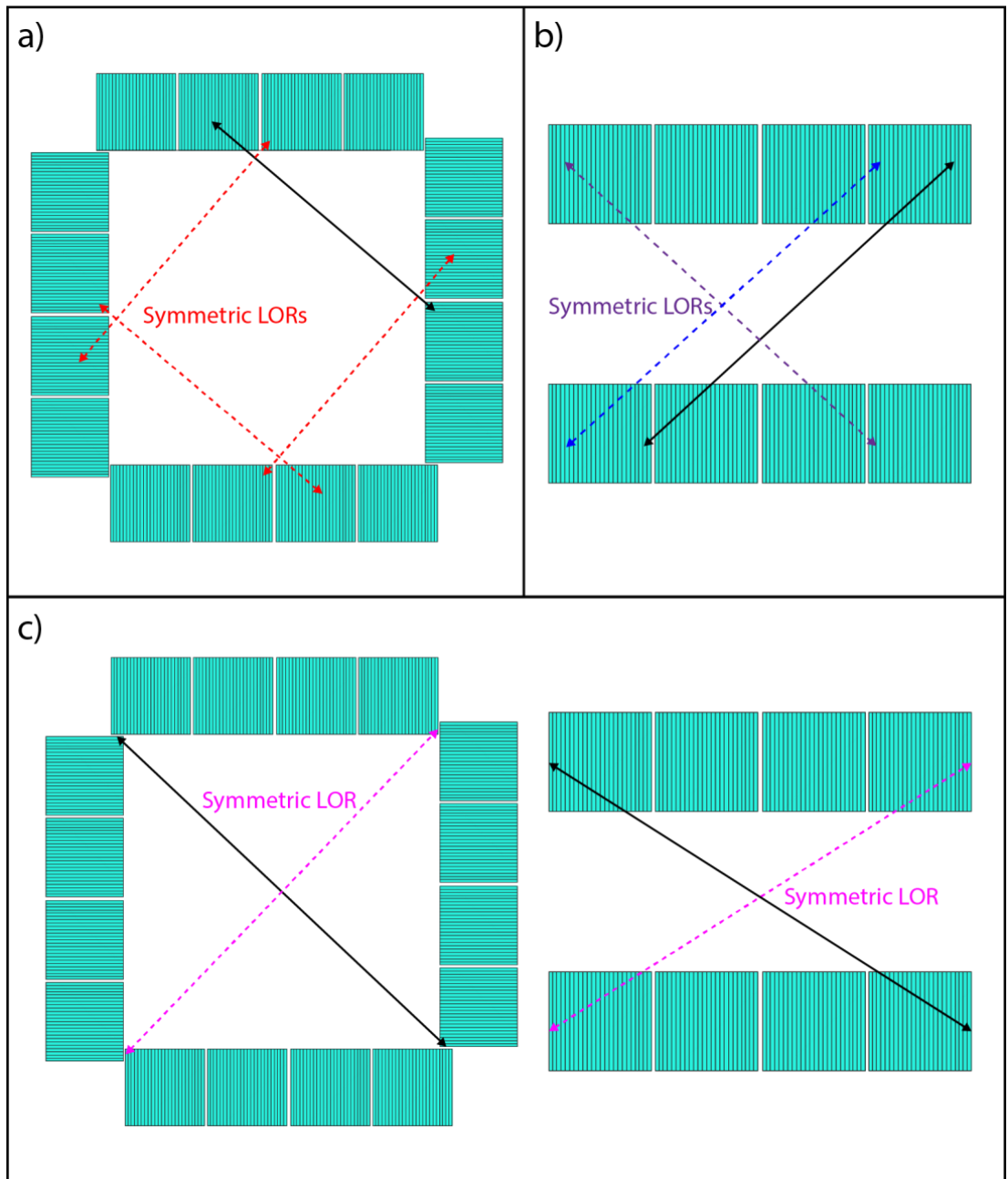

**Figure 3**. Scanner symmetries leveraged to augment count statistics per LOR before implementing normalisation: (**A**) three symmetries in the transaxial plane (dashed red lines) by rotating each LOR in 90-degree increments; (**B**) four more symmetries by mirroring the LOR across the axial plane (1 symmetry, purple dashed line) and rotating the mirrored LOR on the transaxial plane (3 symmetries, blue dashed line as one example); (**C**) the subset of LORs exhibiting no more than three symmetries (dashed magenta lines). The third symmetry (not shown) is derived by a 90-degree rotation of the mirrored LOR (bottom right, dashed magenta line) along the axial plane.

2.4. Simulation Experiments
2.4.1. Spatial Resolution Measurements

To compare the reconstruction performance of the physical scanner and the virtual PET implementations, we performed simulations using a fluorine-18 point source of radius 0.0625 mm. The source was located within the central axial plane of a cylindrical water phantom covering the entire FOV, and offset transaxially to 4 positions: 0, 10, 20, and 25 mm from the centre (Fig. 4a). We applied a point source to background ratio of 1.1:1 (i.e., the point source was 10% hotter than the background) to avoid artificial improvement of the system resolution when using iterative reconstruction (Gong et al 2016). After subtracting the background from the reconstructed source, the resulting point spread function (PSF) was fitted using a double Gaussian. Full-width-at-half-maximum (FWHM) and full-width-at-tenth-maximum (FWTM) values were estimated along four directions: radial, tangential, axial and diagonal (Fig. 4a). In all cases, 200 iterations of MLEM (i.e. 1 subset) were performed before fitting the data.

Volumetric PSF values were measured from the normalised reconstructed PSFs. The volumes were calculated from the number of voxels exceeding a specific threshold (0.5 for FWHM and 0.1 for FWTM). To ensure comparability with measurements from conventional PET scanners, we incorporated a correction factor of $6/\pi$ before presenting the results. The correction factor accounts for the volume disparity between an ellipsoid (representing a 3D PSF) and the circumscribed box resulting from the product of the radial, tangential and axial PSF values.

2.4.2. Micro Derenzo Phantom
A micro Derenzo phantom filled with fluorine-18 was simulated in GATE to evaluate the influence of the system PSF on the resolution of high-frequency objects, while also facilitating a comparative analysis of the reconstructed images from the mousePET scanner and virtual PET implementations. To expedite the simulation and reconstruction processes, we focused on the 0.7 mm, 0.9 mm and 1.2 mm hot-rod diameters only (Fig. 4b). The number of coincidence events during reconstruction was set to approximately 70 million events in all cases, and 10 subsets and 10 iterations were performed for all geometries. For the 6-bin DOI PET geometries, a 2D image-based resolution modelling method (Angelis et al 2013) was applied to assess the effect of PSF modelling on image quality. This was done by performing 5 full OSEM iterations (10 subsets) in which each sub-iteration was interleaved with 2 iterations of a Lucy-Richardson deconvolution algorithm. For each geometry, the reconstructed PSF at the centre of the FOV (section 2.4.1) was used as the deconvolution kernel.

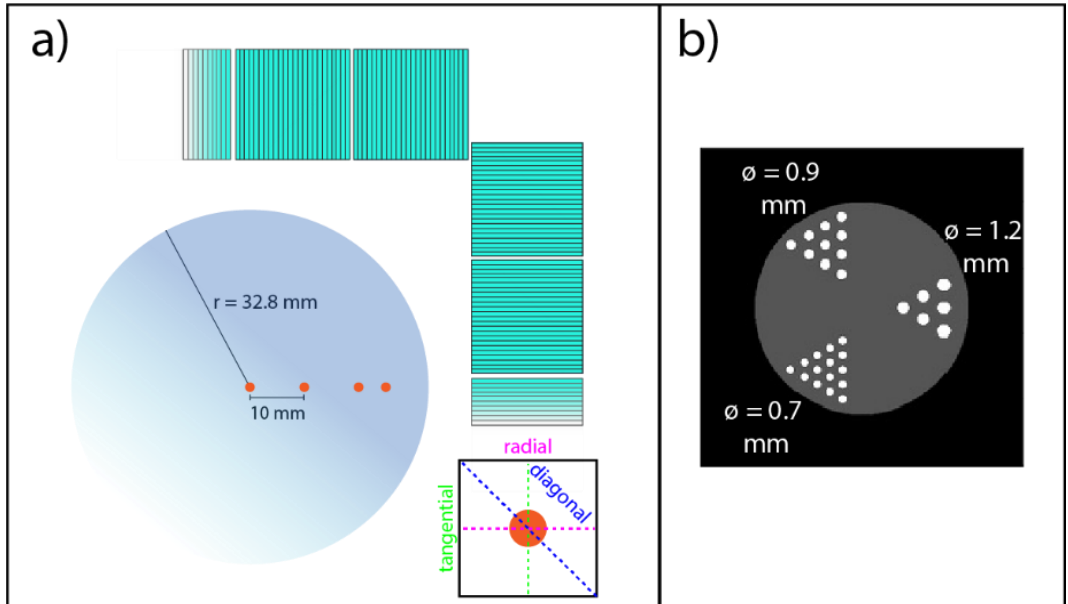

**Figure 4**. (**A**) Fluorine-18 point sources (red circles) placed at the centre of the axial FOV and at different transaxial offsets. Dashed lines indicate the direction of the PSF measurements, while axial profiles (not shown) were along the *z* axis. (**B**) Transaxial view of the 3 hot-rod segments of a simulated micro Derenzo phantom (∅ = 0.7, 0.9 and 1.2 mm).

## 3. Results

3.1. PSF Measurements

Figures 5 and 6 show the zoomed reconstructed PSFs for the physical scanner (mousePET) and virtual PET implementations with the point source offset radially by 0 mm (Fig. 5) and 20 mm (Fig. 6) in the central transaxial plane. Improved DOI resolution (top to bottom row) results in a clear reduction in the characteristic tails of the PSF caused by the rectangular scanner geometry (Qi et al 2002, Kyme et al 2017). Figure 7 shows FWHM along the radial, tangential, axial and diagonal directions. Several features are notable: (i) there was a modest (approximately 8%) improvement in FWHM in transitioning from 4 to 6 DOI bins for the radial component across all geometries; (ii) additional DOI bins reduced the deterioration of the tangential profile of the offset point source; (iii) for a given number of DOI bins, the largest difference in FWHM between the physical scanner and virtual implementations was in the axial and diagonal profiles. Overall, the physical scanner exhibited the highest resolution, followed in order by vPET3, vPET2 and vPET1. For example, for the 6-bin DOI geometries and the source located at a radial offset of 10 mm, the measured diagonal FWHM values were 0.740, 0.763, 0.804 and 0.851 mm for mousePET, vPET3, vPET2 and vPET1, respectively.

The FWTM data (Figure 8) show a reduction in the tails by a factor of 3 and 4 when 4-bin and 6-bin DOI, respectively, was incorporated during reconstruction. Similarly to

FWHM, the main difference in FWTM between the physical scanner and virtual implementations (for the same number of DOI bins) was in the axial and diagonal profiles.

PSF measurements for a fifth point source position, offset by ¼ of the axial FOV on the central axis, were also analysed and these data are included in the supplementary material (Fig. S1).

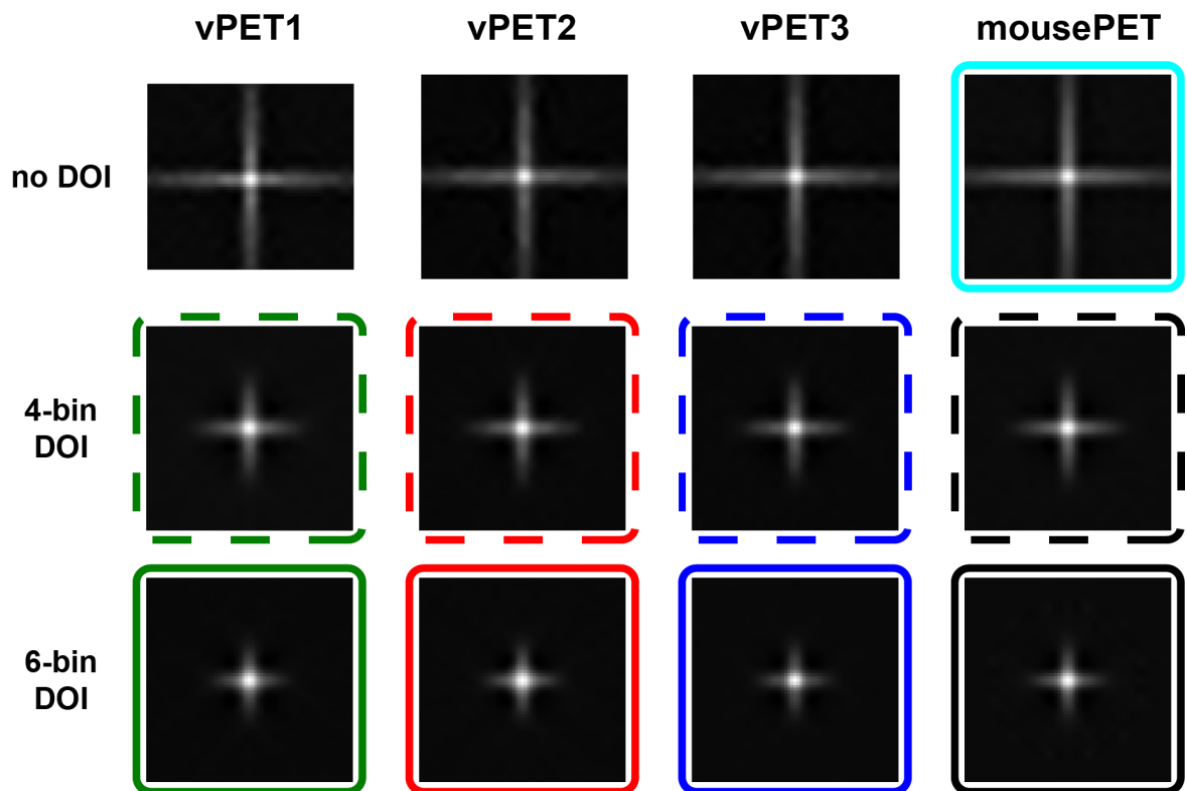

**Figure 5**. Zoomed PSF (after background subtraction) for a point source at the centre of the FOV, shown for the physical scanner and virtual PET implementations (columns) with 3 DOI resolutions (rows).

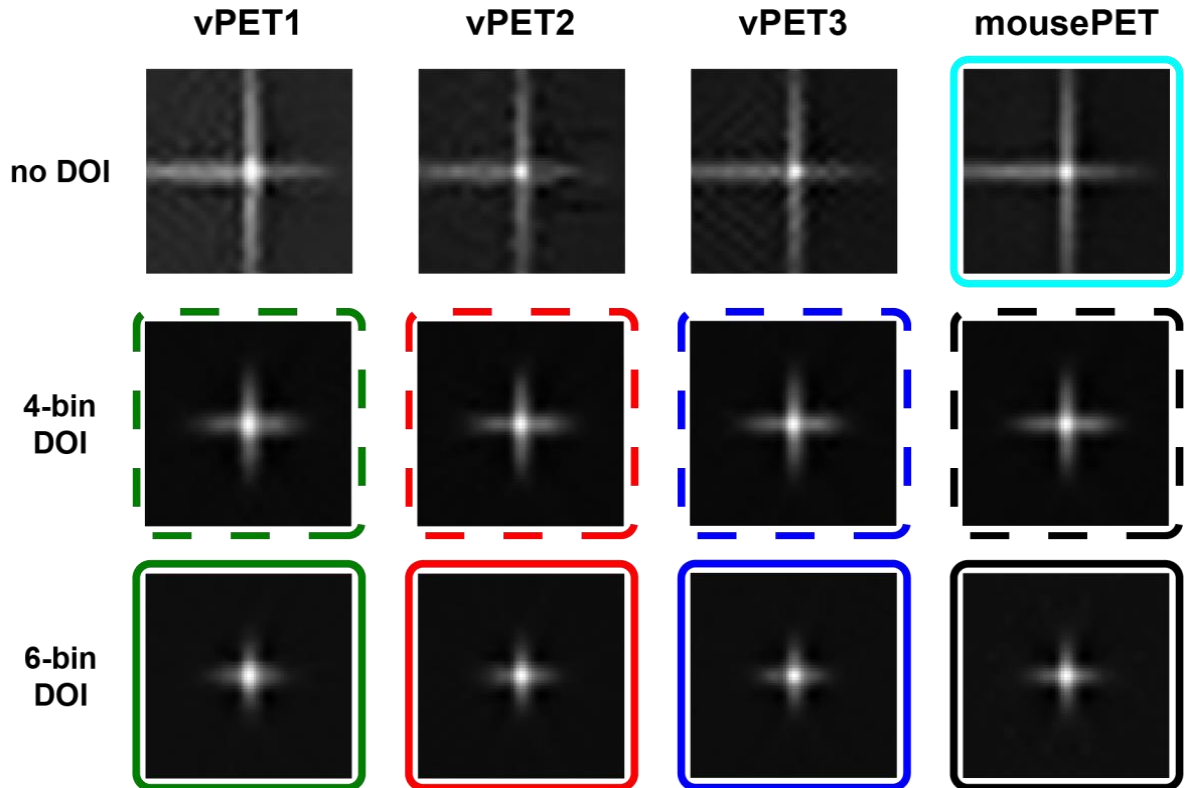

**Figure 6**. Zoomed PSFs (after background subtraction) for a source 20-mm radially offset in the central transaxial plane.

Figure 9 illustrates the volumetric FWHM (left) and FWTM (right) values for the physical scanner and virtual PET implementations with 4-bin (dashed lines) and 6-bin (unbroken lines) DOI, respectively. The unbroken cyan line corresponds to measurements for the non-DOI physical scanner. Overall, at the FWHM level, the improvements for the virtual implementations when transitioning from 4 to 6 DOI bins were 8.88%, 7.06% and 12.17% for vPET 1, 2 and 3, respectively. For the physical scanner the improvement was 14.73%.

3.2. Micro Derenzo Reconstructions

Figure 10 shows the reconstructed central slice of the micro Derenzo phantom for the different geometries as a function of the number of DOI bins used during reconstruction. Despite sub-millimetre FWHM values for the non-DOI physical scanner (Figure 7), the long PSF tails led to a very poor reconstruction with no rods being clearly distinguishable. In contrast, using 4-bin or 6-bin DOI resulted in the 0.9- and 1.2-mm rods being resolvable for all geometries. Figure 11 compares line profiles through the 0.9- and 1.2-mm rods (top) and through the 0.9- and 0.7-mm rods (bottom). The 4- to 6-bin DOI transition resulted in improved contrast recovery coefficients for almost all rods (expected activity: 55 a.u.). Tables 2 and 3 show the peak-to-valley ratios for the different rods and for the 4- and 6-bin DOI geometries, respectively. A mean improvement of 25% was obtained when transitioning from 4 to 6 DOI bins.

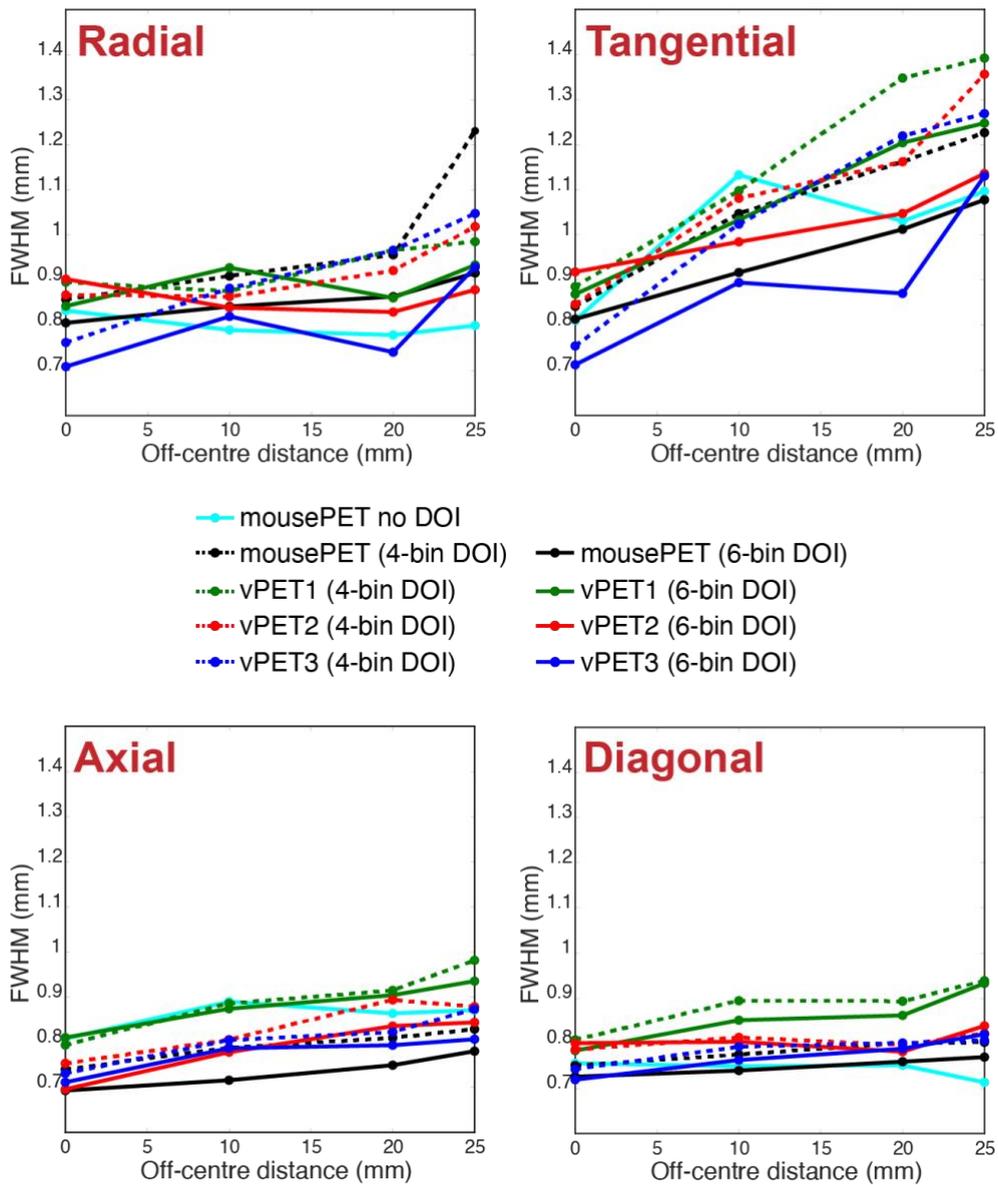

**Figure 7**. Radial, tangential, axial and diagonal FWHM measurements for the physical scanner and virtual PET implementations with 3 DOI resolutions. Dashed and unbroken lines correspond to measurements performed using 4 and 6 DOI bins during reconstruction, respectively. In each case the cyan unbroken line represents the performance for the non-DOI physical scanner.

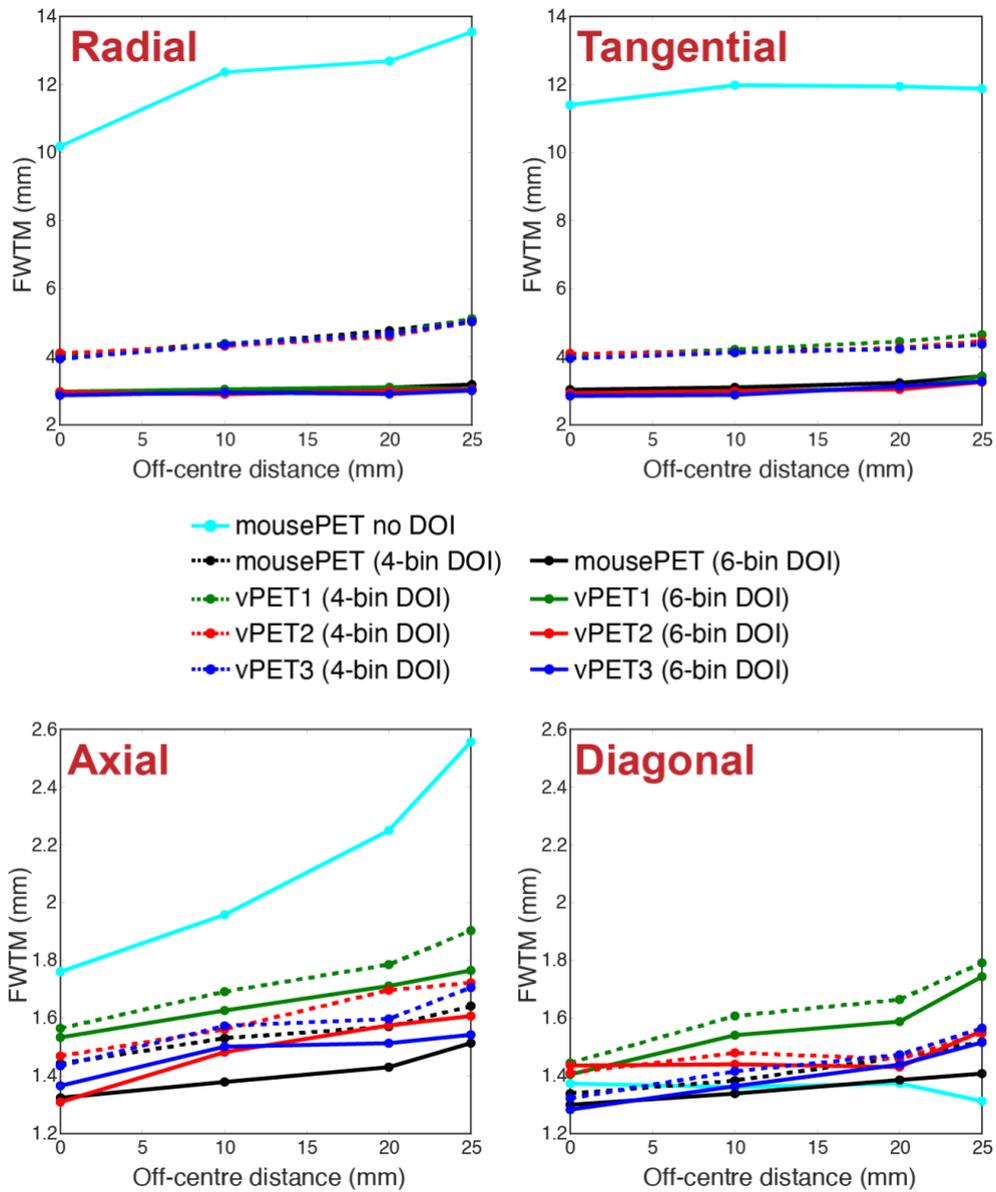

**Figure 8**. Radial, tangential, axial and diagonal FWTM measurements for the physical scanner and virtual PET implementations with 3 DOI resolutions. Line colours and styles are the same as in figure 7.

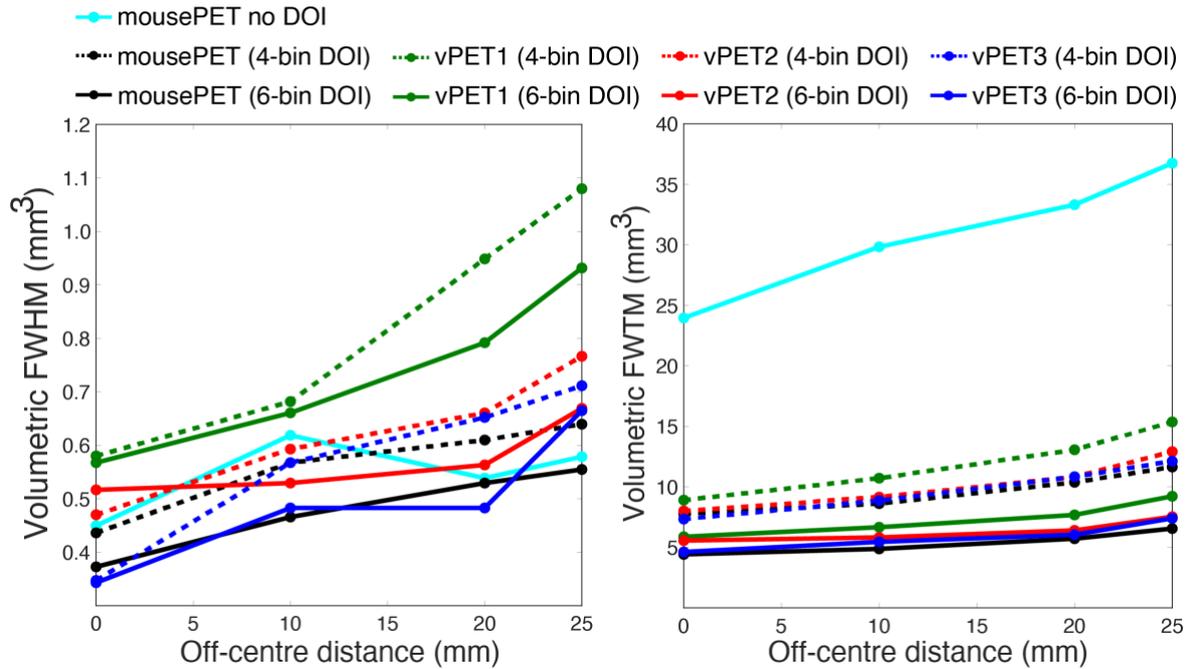

**Figure 9**. Volumetric FWHM (left) and FWTM (right) for the physical scanner and virtual PET implementations with 3 DOI resolutions. Line colours and styles are the same as in figure 7.

Figure 12 compares the micro Derenzo reconstructions with and without PSF modelling. PSF modelling resulted in suppression of the PSF tails, which was more pronounced for the larger rods. It provided no discernible improvement in the resolvability of the 0.7-mm rods. For all geometries, the improvement in the peak-to-valley ratio due to PSF modelling was: 12%, 62% and 145% for the 0.7-mm, 0.9-mm, and 1.2-mm rods, respectively. A direct comparison between the non-PSF and PSF-modelled OSEM reconstructions is provided in the supplementary data (Fig. S2).

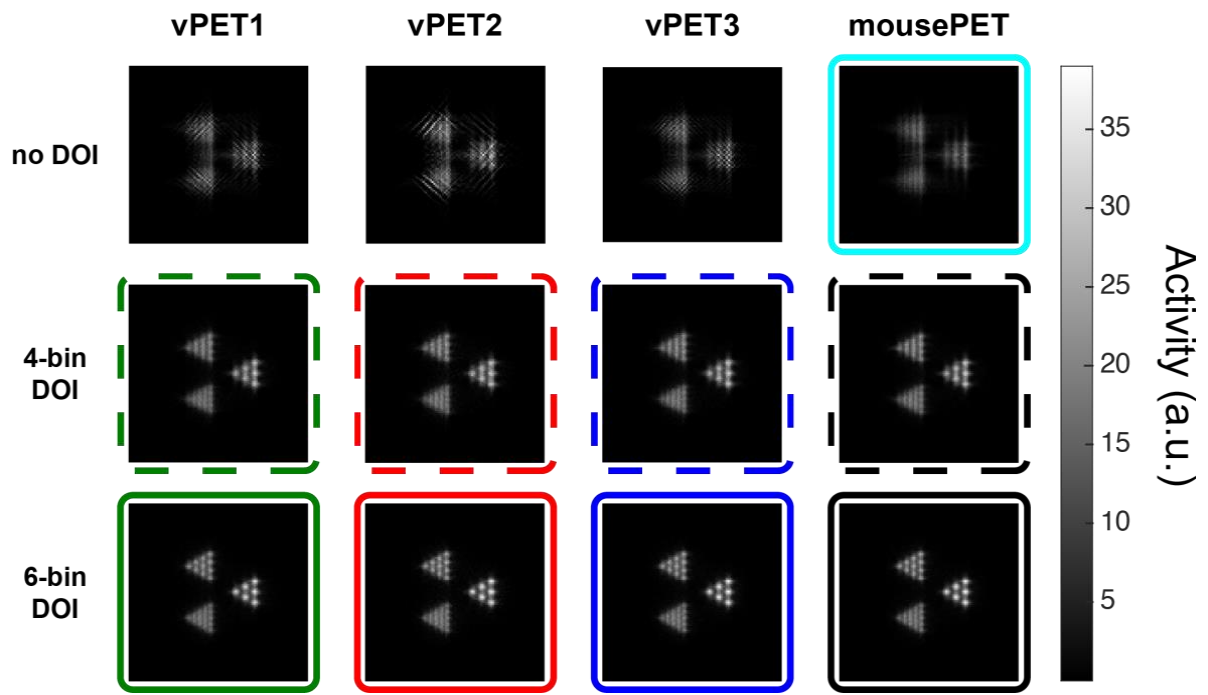

**Figure 10.** Micro Derenzo reconstructions (central slice) for the physical scanner and virtual PET implementations (columns) with different DOI resolution (rows).

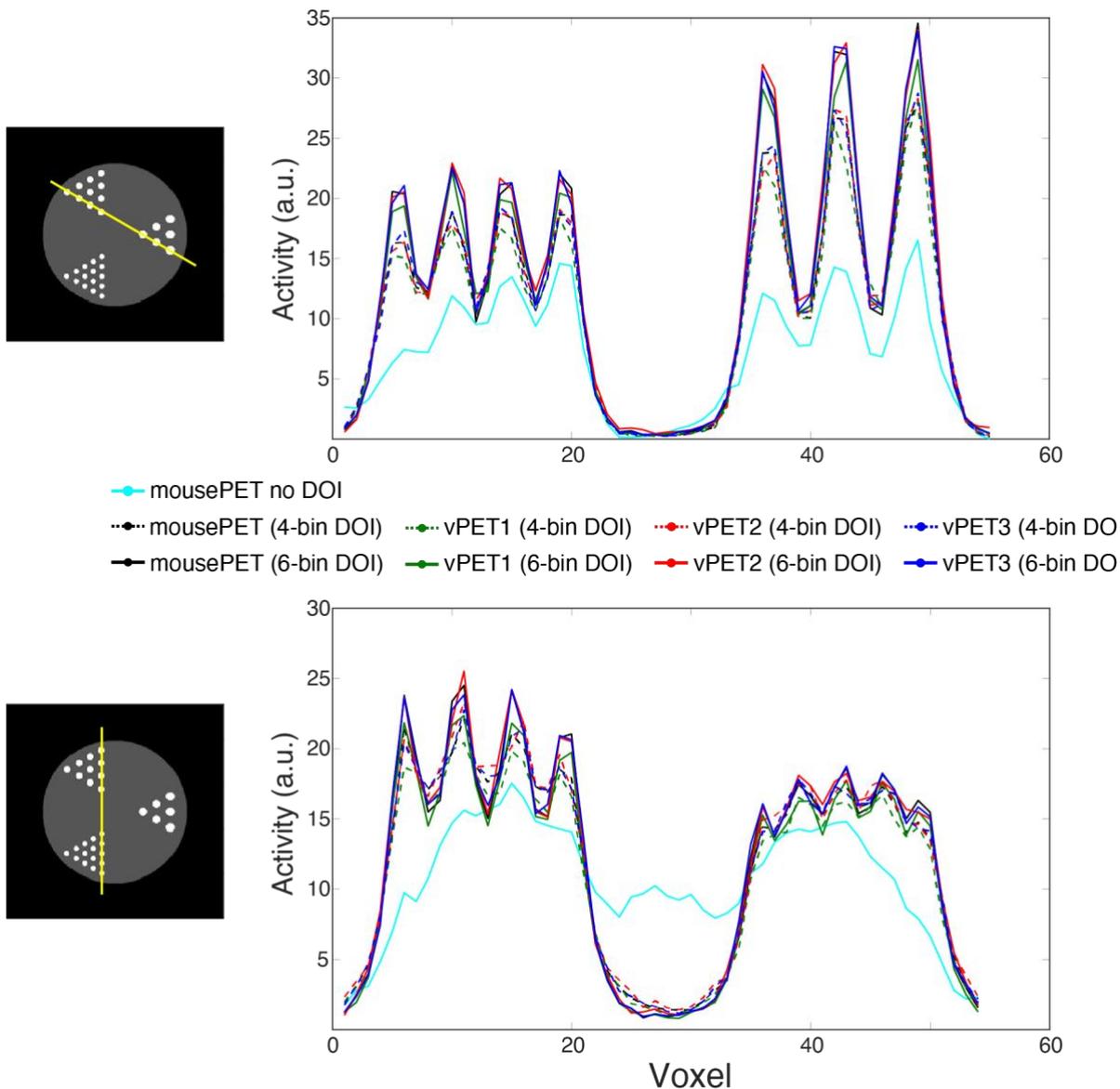

**Figure 11.** Line profiles through the 0.9 and 1.2 mm rods (**top**), and the 0.9 and 0.7 mm rods (**bottom**) in the reconstructions of the micro Derenzo phantom, shown for different DOI resolutions.

Table 2. Peak-to-valley ratios across the 0.7-mm, 0.9-mm and 1.2-mm rods for the 4-bin DOI geometries.

| Rod Diameter (mm) | vPET1 | vPET2 | vPET3 | mousePET |
|---|---|---|---|---|
| 0.7 | 0.86 | 0.58 | 0.81 | 1.12 |
| 0.9 | 1.55 | 1.52 | 1.65 | 1.62 |
| 1.2 | 2.34 | 2.31 | 2.45 | 2.37 |

Table 3. Peak-to-valley ratios across the 0.7-mm, 0.9-mm and 1.2-mm rods for the 6-bin DOI geometries.

| Rod Diameter (mm) | vPET1 | vPET2 | vPET3 | mousePET |
|---|---|---|---|---|
| 0.7 | 1.19 | 0.85 | 1.18 | 1.18 |
| 0.9 | 1.75 | 1.90 | 1.92 | 1.98 |
| 1.2 | 2.86 | 2.84 | 2.91 | 3.01 |

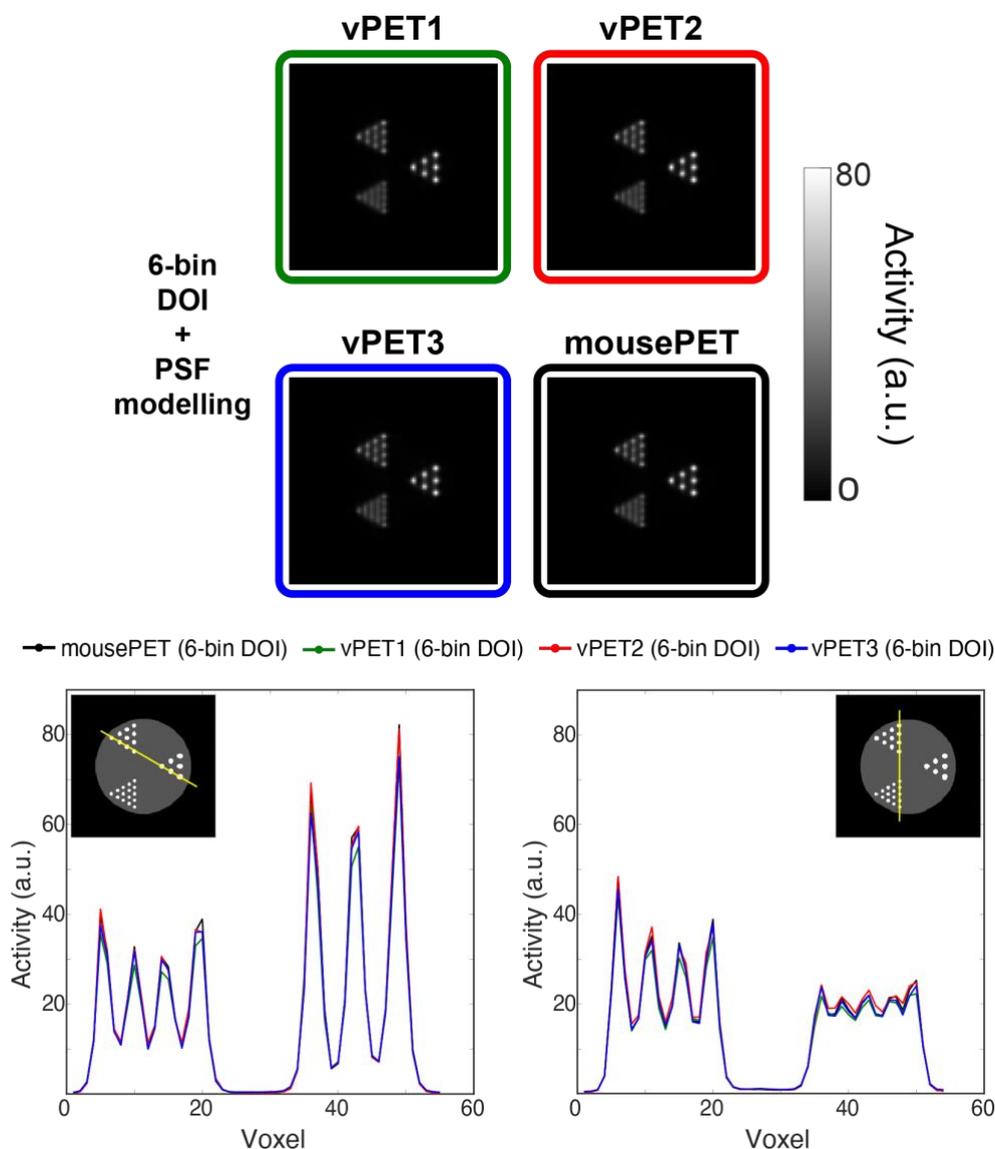

**Figure 12.** (**top & middle**) Micro Derenzo phantom reconstructions with 6-bin DOI resolution and PSF modelling. (**bottom**) Line profiles through the 0.9 and 1.2 mm rods (left) and the 0.9 and 0.7 mm rods (right).

## 4. Discussion

This study investigated the impact of virtual detector geometries on the spatial resolution/efficiency trade-off, focusing on the open-field mouse brain PET system

(mousePET) as a case study of high resolution and DOI-capable application-specific PET scanners under development. The combination of highly segmented detectors and 3-mm DOI resolution results in almost 7.3 billion LORs for the mousePET system. On-disk storage of the system matrix is impractical considering the number of LORs and voxels (~155 petabytes assuming 4-byte precision), while on-the-fly calculations translate into very lengthy reconstructions. The challenge is exacerbated by the fact that motion compensation with attenuation correction requires repeated calculation of a time-weighted sensitivity matrix (Angelis et al 2014). Extending on previous work, we tested three different virtual cylindrical PET implementations (virtual detector sizes) to investigate the trade-off between LOR data reduction and spatial resolution degradation. For DOI-based reconstruction in mousePET, the virtual implementations provided an LOR reduction of at least 50%, with most achieving resolution values comparable to mousePET.

4.1 Spatial Resolution Measurements

The PSF anisotropy in mousePET, which arises due to the highly non-uniform parallax effect (Kyme et al 2017), makes the radial and tangential profiles more dependent on DOI resolution than on segmentation coarseness of the virtual PET implementations. Instead, most of the resolution degradation introduced by these implementations occurs in the axial and diagonal profiles. The volumetric FWHM and FWTM values provide a practical means of choosing a virtual detector size for an application. The method of calculation we used acknowledges the unusual shape of our asymmetric PSF, which would be unduly penalised if we had used the conventional method of multiplying the 3 independent components (which assumes a symmetric PSF). Based on the volumetric FWHM measurements, it is clear that vPET1 is inferior to the other geometries, while vPET2 and vPET3 have comparable performance except for the central measurement. The difference in the central measurement may stem from variations in the sensitivity matrix of the two systems. Comparison of the central voxel of the sensitivity images showed that for vPET2 (6-bin DOI) this region is poorly sampled. For vPET3 the opposite is true. This effect thus highlights a limitation: different voxel grids may better suit specific virtual geometries. Typically, the voxel size in cylindrical PET systems is chosen as D/2 (or D/3), where D is the detector size. We used a voxel size suitable for the physical scanner but not necessarily optimal for all virtual implementations. Nevertheless, both vPET2 and vPET3 provide almost identical reconstructions to the physical scanner in the transaxial plane. For mousePET we conclude that the optimal choice is a virtual detector size of 0.4712 mm (vPET2 detector size), achieving 86% reduction in LOR data while maintaining sub-millimetre spatial resolution.

4.2. Micro Derenzo Reconstructions

The micro Derenzo phantom reconstructions reveal the impact of the PSF tails when imaging high-frequency objects. Although it is primarily noticeable at the FWTM range, it also affects the peak-to-valley ratio measurements (Table 2). For a given rod set, the ratios depend on the rod arrangement and line profile orientation. We observed that peak-to-valley ratios improved when line profiles were drawn away from the tails (e.g., Fig. 10, top line profile). Ratios shown in Tables 2 and 3 correspond to those from lines not crossing the tails. As an example, for the 1.2-mm rods, deterioration of these ratios when passing through the tails was measured to be 68% and 34% when using 4-bin and 6-bin DOI, respectively. In agreement with the PSF measurements, the best ratios were achieved by the physical scanner followed by the virtual implementations. Although vPET1 achieved better ratios than vPET2, the line profiles indicate that the reconstructed activities in this virtual implementation did not converge to the same reconstructed activity as the other geometries. For the 6-bin DOI geometries, the difference in ratio between vPET2 and the physical scanner was 0.0706 and 0.1765 for the 0.9- and 1.2-mm rods, respectively.

As highlighted previously (Angelis et al 2013), 2D image-based PSF modelling can significantly enhance the reconstructions and increase the recovery coefficient of the hot rods (12-145% improvement in our experiments). Given the unconventional box-shaped geometry of mousePET, resolution modelling is very effective in minimising the effect of the relatively long and asymmetric PSF tails. In future work, the application of PSF modelling using deep learning methods, such as those discussed in (da Costa-Luis and Reader 2017, Song et al 2020), could be beneficial for further resolution improvement of mousePET in the context of virtual detectors.

4.3. Virtual geometries for DOI reconstruction

Although it has been suggested that the classic PET resolution equation (Moses 2011) enables one to calculate the maximum virtual detector size before spatial degradation is observed (Li et al 2015), our results show that this equation does not hold for unconventional geometries. Using this equation, and disregarding positron range and photon acollinearity, we would predict a virtual detector size of 0.8052 mm to give reconstructions with the same resolution as the physical mousePET. Our experimental results show that the Moses equation considerably over-estimates the virtual detector size required to achieve this outcome. Instead, we suggest that equation (2) from (Moses 2011) should only be used as the starting point for optimising the virtual detector size on a particular PET geometry.

Contrary to (Kim et al 2018), our results also indicate that sufficient (sub-millimetre) spatial resolution is preserved without the need for subsampling the image space. Instead, in agreement with (Zhang et al 2016) we note that a simple LOR rebinning (from physical to virtual LOR) is appropriate when the differences between physical and virtual geometries are sufficiently well modelled – here, accounted for through appropriate normalisation of the virtual geometries.

**5. Conclusion**

High resolution, narrow bore PET scanners such as mousePET result in a very large number of LORs due to the highly sampled image space and DOI binning. In such systems, LOR data reduction is desirable due to the computational demands of image reconstruction. The introduction of virtual cylinders for DOI-based reconstruction in such systems can achieve sub-millimetre spatial resolution while reducing the number of LORs by 86%. Preserving high spatial resolution, together with the high level of data compression, makes virtual cylinders an attractive option for application-specific DOI-capable PET scanners. For mousePET and other non-cylindrical geometries, resolution recovery methods are further recommended to better supress the effects of the asymmetric PSF on reconstructed image resolution.

## Acknowledgments

This work was supported by the Australian Research Council (Discovery Projects grant DP190102318).
F.E. Enríquez-Mier-y-Terán is supported by the Mexican National Council of Humanities, Sciences and Technologies (CONAHCYT).